# Ground Canonicity

Nachum Dershowitz*
School of Computer Science
Tel-Aviv University
Ramat Aviv, Tel-Aviv 69978, Israel
email: Nachumd@tau.ac.il

April 2003

> *They are not capable
> to ground a canonicity
> of universal consistency.*
>
> —Alexandra Deligiorgi
> (ΠΑΙΔΕΙΑ, 1998)

**Abstract**

We explore how different proof orderings induce different notions of saturation. We relate completion, paramodulation, saturation, redundancy elimination, and rewrite system reduction to proof orderings.

## 1 Introduction

We show how to define the *canonical* basis of an abstract deductive system in three distinct ways: (1) Formulæ appearing in minimal proofs; (2) non-redundant lemmata; (3) minimal trivial theorems. Well-founded orderings of proofs [Bachmair and Dershowitz, 1994] are used to distinguish between cheap "direct" proofs, those that are of a computational flavor (e.g. rewrite proofs), and expensive "indirect" proofs, those that require search to find. This approach suggests generalizations of the concepts of "redundancy" and "saturation", as elaborated by Nieuwenhuis and Rubio in [2001]. Saturated, for us, means that all *cheap* proofs are supported. By considering different orderings on proofs, one gets different kinds of saturated sets.

This work continues our development of an abstract theory of "canonical inference", initiated in [Dershowitz and Kirchner, 2003b]. Although we will use ground equations as an illustrative example, the framework applies equally well in the first-order setting, whether equational or clausal. Though our motivation is primarily æsthetic; our expectation is that practical applications will follow.

---

*Supported in part by the Israel Science Foundation.

Proofs and other details omitted here will be included in [Dershowitz and Kirchner, 2003a]. Our rewriting terminology accords with [Dershowitz and Plaisted, 2001].

## 2 Proof Systems

Let $\mathbb{A}$ be the set of all formulæ (ground equations and disequations, in our examples) over some fixed vocabulary. Let $\mathbb{P}$ be the set of all (ground equational) proofs. We are given two functions: $\Gamma : \mathbb{P} \to 2^{\mathbb{A}}$ gives the assumptions in a proof, and $\Delta : \mathbb{P} \to \mathbb{A}$ gives its conclusion. Both are extended to sets of proofs in the usual fashion. (We assume for simplicity that proofs use only a finite number of assumptions.)

The framework proposed here is predicated on two *well-founded* partial orderings over $\mathbb{P}$: a *proof ordering* $\geq$; and a *subproof relation* $\trianglerighteq$. They are related by a monotonicity requirement given below (7).

We will use the term *presentation* to mean a set of formulæ, and *justification* to mean a set of proofs. We reserve the term *theory* for deductively closed presentations. Let $A^*$ denote the *theory* of presentation $A$, that is, the set of conclusions of all proofs with assumptions $A$:

$$A^* \quad := \quad \Delta\,\Gamma^{-1}A \;=\; \{\Delta\,p : p \in \mathbb{P},\ \Gamma\,p = A\} \tag{1}$$

We assume the following three standard properties of Tarskian consequence relations:

$$A^* \;\subseteq\; (A \cup B)^* \tag{2}$$
$$A \;\subseteq\; A^* \tag{3}$$
$$A^{**} \;=\; A^* \tag{4}$$

Thus, $-^*$ is a closure operation. We say that presentation $A$ is a *basis* for theory $B$ if $A^* = B$. Presentations $A$ and $B$ are *equivalent* if their theories are identical: $A^* = B^*$.

As a very simple running example, let the vocabulary consist of the constant 0 and unary symbol $s$. Abbreviate tally terms $s^i 0$ as numeral $i$. The set $\mathbb{A}$ consists of all *unordered* equations $i = j$ (so symmetry is built into the structure of proofs). We postpone dealing with disequations for the time being. An equational inference system for this vocabulary might consist of the following inference rules:

$$\frac{\Box}{0 = 0}\ \mathbf{Z} \qquad \frac{\boxed{i = j}}{i = j}\ \mathbf{I}_{i=j}$$

$$\frac{i = j \quad j = k}{i = k}\ \mathbf{T} \qquad \frac{i = j}{si = sj}\ \mathbf{S}$$

where the proof tree branches (of $T$) are *unordered*. To accommodate (2), we also need projection:

$$\frac{a \quad c}{c}\ \mathbf{P}$$

For example, if $A = \{4 = 2, 4 = 0\}$, then $A^* = \{i = j : i \equiv j \pmod{2}\}$.

Consider the proof schemata:

$$
\begin{array}{c}
\dfrac{\Box}{0=0} \\
\dfrac{}{1=1} \\
\vdots \\
\hline i=i
\end{array}
\qquad
\dfrac{\dfrac{\boxed{4=0}}{4=0}\quad \dfrac{\boxed{4=2}}{4=2}}{\dfrac{2=0}{\dfrac{i-j=0}{\vdots\atop i=j}}}
\qquad
\dfrac{\dfrac{\vdots}{i-j-2=0}}{\dfrac{i-j-1=1}{i-j=2}}
$$

Let's use proof terms for proofs, denoting the above two trees by $S^i Z$ and $S^j T(T(I(4,0), I(4,2)), SS(\nabla_{i-j-2=0}))$, respectively. With a recursive path ordering to order proofs, precedence $Z < S < T < I < P < 0 < 1 < 2 < \cdots$, and multiset "status" for $I$, minimal proofs of the theorems in $A^*$ must take one of these two forms, or the form of one of their subproofs.

We call a proof *trivial* when it proves only itself and has no subproofs other than itself, that is, if $\Gamma p = \{\Delta p\}$ and $p \trianglerighteq q \Rightarrow p = q$. We denote by $\widehat{a}$ such a trivial proof of $a \in \mathbb{A}$ and by $\widehat{A}$ the set of trivial proofs of each $a \in A$. We assume that proofs use their assumptions, that subproofs don't use non-existent assumptions, and—most significantly—that proof orderings are monotonic with respect to subproofs. Specifically, for all proofs $p, q, r$ and formulæ $a$:

$$a \in \Gamma p \;\Rightarrow\; p \trianglerighteq \widehat{a} \tag{5}$$
$$p \trianglerighteq q \;\Rightarrow\; \Gamma p \supseteq \Gamma q \tag{6}$$
$$p \triangleright q > r \wedge \Delta q = \Delta r \;\Rightarrow\; \exists v \in \mathbb{P}.\; p > v \triangleright r \wedge \Delta p = \Delta v \tag{7}$$

We make no other assumptions regarding proofs or their structure.

Every formula $a$ admits a trivial proof $\widehat{a}$ by (3,5). Let $\Sigma p = \{q \colon p \trianglerighteq q\}$ denote the subproofs of $p$, and likewise $\Sigma P = \cup_{p \in P} \Sigma p$. This way, (5) can be abbreviated $\widehat{\Gamma p} \subseteq \Sigma p$.

It may be convenient to think of a proof-tree "leaf" as a subproof with only itself as a subproof; other subproofs are the "internal nodes". There are two kinds of leaves: trivial proofs $\widehat{a}$ (such as inferences **I**), and vacuous proofs $\bar{a}$ with $\Gamma \bar{a} = \emptyset$ and $\Delta \bar{a} = a$ (such as **Z**). By well-foundedness of $\trianglerighteq$, there are no infinite "paths" in proof trees.

Postulate (7) states that $>$ (restricted to proofs with the same conclusion) and $\triangleright$ commute (i.e. $\triangleright \circ > \;\subseteq\; > \circ \triangleright$), from which it follows that their union, the partial ordering $(\triangleright \cup >)^*$, is also well-founded.

## 3 Canonical Systems

Denote the set of proofs using assumptions $A$ by:

$$\Pi A \quad := \quad \{p \in \mathbb{P} \colon \Gamma p \subseteq A\}$$

and define the *minimal* proofs in a set of proofs as:

$$\mu P \quad := \quad \{p \in P : \neg \exists q \in P.\, \Delta\, q = \Delta\, p,\ q < p\}$$

On account of well-foundedness, minimal proofs always exist.

**Proposition 1** *For all presentations $A, B$:*

$$\Gamma \Pi A \subseteq A \tag{8}$$
$$\Sigma \mu \Pi A \subseteq \mu \Pi A \tag{9}$$
$$\Pi \Gamma \Pi A = \Pi A \tag{10}$$
$$\Pi A = \Pi B \Leftrightarrow A = B \tag{11}$$

*And for all justifications $P$: $P \subseteq \Pi \Gamma P$.*

We say that presentation $A$ is *reduced* when $A = \Gamma\, \mu \Pi A$. Our main definition is:

**Definition 1 (Canonical Presentation)** *The* canonical presentation *contains those formulæ that appear as assumptions of minimal proofs:*

$$A^\sharp \quad := \quad \Gamma\, \mu \Pi A^*$$

*So, we say that $A$ is* canonical *if $A = A^\sharp$.*

Proof orderings are lifted to sets of proofs, as follows:

**Definition 2** *Justification $Q$ is* better *than justification $P$ if:*

$$P \sqsupseteq Q \quad :\equiv \quad \forall p \in P.\, \exists q \in Q.\, \Delta\, q = \Delta\, p \wedge p \geq q$$

*It is* much better *if:*

$$P \sqsupset Q \quad :\equiv \quad \forall p \in P.\, \exists q \in Q.\, \Delta\, q = \Delta\, p \wedge p > q$$

*Justifications are* similar *if:*

$$P \simeq Q \quad :\equiv \quad P \sqsupseteq Q \sqsupseteq P$$

These three relations are compatible: $\sqsupseteq \circ \sqsupset \subseteq \sqsupset$, $\sqsupseteq \circ \simeq \subseteq \sqsupseteq$, etc.

The following statements can be shown to hold:

**Proposition 2** *For all justifications $P, Q$:*

$$P \sqsupseteq \mu P \tag{12}$$
$$P \sqsupseteq Q \Leftrightarrow \mu P \sqsupseteq \mu Q \tag{13}$$
$$P \sqsupset Q \Leftrightarrow \mu P \sqsupset \mu Q \tag{14}$$
$$P \simeq Q \Leftrightarrow \mu P = \mu Q \tag{15}$$

**Proposition 3** *The relation $\sqsupseteq$ is a quasi-ordering on proofs and a partial ordering of* minimal *proofs.*

This quasi-ordering on proofs is lifted to sets of formulæ as follows:

**Definition 3** *Presentation $B$ is said to be* simpler *than an equivalent presentation $A$ when $B$ provides better proofs than does $A$: $A \succsim B \quad :\equiv \quad A^* = B^* \wedge \Pi A \sqsupseteq \Pi B$. Presentations are* similar *if:* $A \approx B \quad :\equiv \quad \Pi A \simeq \Pi B$.

These relations are also compatible.

**Proposition 4** *For all presentations $A, B$:*

$$\Pi A \quad \sqsupseteq \quad \Pi(A \cup B) \tag{16}$$
$$A \approx B \quad \Leftrightarrow \quad \mu\Pi A = \mu\Pi B \tag{17}$$
$$B \subseteq A \wedge \Pi A \sqsupseteq \Pi B \quad \Rightarrow \quad A \approx B \tag{18}$$
$$A \subseteq B \wedge A^* = B^* \quad \Rightarrow \quad A \succsim B \tag{19}$$

**Proposition 5** *The relation $\succsim$ is a quasi-ordering and $\approx$ is its associated equivalence relation.*

The function $\_^\sharp$ is "canonical" with respect to equivalence of presentations. That is: $A^{\sharp *} = A^*$; $A^* = B^* \Leftrightarrow A^\sharp = B^\sharp$; and $A^{\sharp \sharp} = A^\sharp$. This justifies the terminology of Definition 1.

**Lemma 1** $A \succsim A^\sharp$.

## 4  Saturated Systems

By a "normal-form proof", we will mean a proof in $\mu\Pi A^*$. On account of (7), all subproofs of normal-form proofs are also in normal form. We propose the following definitions:

**Definition 4 (Saturation)** *A presentation $A$ is* saturated *if it supports all possible normal form proofs: $\Pi A \supseteq \mu\Pi A^*$. A presentation $A$ is* complete *if every theorem has a normal form proof: $A^* = \Delta (\Pi A \cap \mu\Pi A^*)$.*

A presentation is complete if it is saturated, but for the converse, we need a further hypothesis: *minimal proofs are unique* if for all theorems $c \in \Pi A$ there is exactly one minimal proof in $\mu\Pi A^*$ with conclusion $c$.

**Proposition 6** *If minimal proofs are unique, then a presentation is saturated iff it is complete.*

For example, suppose all rewrite (valley) proofs are minimal but incomparable. Then any Church-Rosser system is complete, since every identity has a rewrite prof, but only the full deductive closure is saturated.

**Theorem 1** *A presentation $A$ is saturated iff it contains its own canonical presentation: $A \supseteq A^\sharp$. In particular, $A^\sharp$ is saturated. Moreover, $A^\sharp$ is the smallest saturated set: no equivalent proper subset of $A^\sharp$ is saturated; if $A$ is saturated, then every equivalent superset also is.*

**Corollary 1** *Presentation $A$ is saturated iff $A^* \approx A$.*

**Proof.** It is always the case that $A \succsim A^* \succsim A^\sharp$. If $A$ is saturated, then $A \supseteq A^\sharp$ and, therefore, $A^* \succsim A^\sharp \succsim A$. For the other direction, suppose $p \in \mu\Pi A^*$. Since $A$ is similar, there must be a proof $q \in \Pi A \subseteq \Pi A^*$, such that $q \leq p$. But $q \not< p$, so $p \in \Pi A$. It follows that $\mu\Pi A^* \subseteq \Pi A$, and $A$ is saturated.

**Lemma 2** *Similar presentations are either both saturated or neither is; similar presentations are either both complete or neither is.*

**Proof.** The first claim follows directly from the previous result. For the second, one can verify that $A \approx B$ implies:

$$\begin{aligned} B^* &= A^* = \Delta\left(\Pi A \cap \mu\Pi A^*\right) = \Delta\left(\mu\Pi A \cap \mu\Pi A^*\right) \\ &= \Delta\left(\mu\Pi B \cap \mu\Pi B^*\right) = \Delta\left(\Pi B \cap \mu\Pi B^*\right) \end{aligned}$$

Formulæ that can be removed from a presentation—without making proofs worse—are "redundant":

**Definition 5 (Redundancy)** *A set $R$ of formulæ is* (globally) redundant *with respect to a presentation $A$ when: $A \cup R \succsim A \setminus R$. The set of all* (locally) redundant *formulæ of a given presentation $A$ will be denoted $\rho A$:*

$$\rho A \quad := \quad \{r \in A \colon A \succsim A \setminus \{r\}\}$$

*A presentation $A$ is* irredundant *if $\rho A = \emptyset$.*

It can be shown that $A$ is reduced iff it is irredundant.

**Lemma 3** *The canonical presentation is reduced: $\rho A^\sharp = \emptyset$.*

**Proposition 7** *The following facts hold for all presentations $A$:*

$$\begin{aligned} A &\approx A \setminus \rho A & (20) \\ A^\sharp &= A^* \setminus \rho A^* & (21) \\ A^\sharp &= \Delta(\mu\Pi A^* \cap \widehat{A^*}) & (22) \\ \widehat{A^\sharp} &= \mu\Pi A^* \cap \widehat{A^*} & (23) \end{aligned}$$

It is thanks to well-foundedness of $>$ that the set of all *locally* redundant formulæ in $\rho A$ is *globally* redundant (Eq. 20). The alternate definition of the canonical set (22) is made possible by the properties of subproofs.

**Theorem 2** *A presentation is canonical iff it is saturated and reduced.*

**Proof.** One direction follows immediately from Theorem 1 and Lemma 3. For the other direction, let $A$ be saturated and reduced. We aim to show that $A = A^\sharp$. By Lemma 1, $A \succsim A^\sharp$ and the two presentations are equivalent. If $A$ is saturated, then by Theorem 1, $A \supseteq A^\sharp$. By (19), for any $r \in A \setminus A^\sharp$, $A \succsim A^\sharp \succsim A \setminus \{r\}$. But $\rho A = \emptyset$, since $A$ is reduced, so it cannot be that $r \in A$. In other words, $A \setminus A^\sharp = \emptyset$, and $A$ is canonical.

Returning to our simple example, we can add three inference rules for disequalities:

$$\frac{i = j \quad j \neq k}{i \neq k} \mathbf{T} \qquad \frac{i \neq i}{j = k} \mathbf{F}_{j=k} \qquad \frac{\boxed{i \neq j}}{i \neq j} \mathbf{I}_{i \neq j}$$

With them, one can infer, for example, $0 \neq 0$ from $1 \neq 1$. If $F$ is smaller than other proof combinators, and $I$ nodes are incomparable, then the canonical basis of any inconsistent set is $\{i \neq j : i, j \in \mathbf{N}\}$. All positive equations are redundant.

## 5 Variations

Consider the above inference rules for ground equality and disequality: $S, T, F, I, Z$, with $S$ extended to apply to all function symbols of any arity. Suppose we are using something like the recursive path ordering for proof terms.

**Refutation.** If the inference rule $F$ is the cheapest in the proof ordering, $T < I$, and $I(i, j)$ nodes are measured by the values of $i$ and $j$, then the canonical basis of any inconsistent presentation is a (smallest) trivial disequation $\{t \neq t\}$.

**Deduction.** If the proof ordering prefers direct application $I$ of axioms over all other inferences (including $Z$), then trivial proofs are best. In that case, $\rho A^* = \emptyset$ and the canonical basis includes the whole theory $A^\sharp = A^*$.

**Paramodulation.** If the proof ordering makes functional reflexivity $S$ smaller than $I$, but the only ordering on leaves is $I(u, t) \leq I(c[u], c[t])$ for any context $c$, then the canonical basis will be the congruence closure, as generated by paramodulation: $\rho A = \{f(u_1, \ldots, u_n) = f(t_1, \ldots, t_n) : u_1 = t_1, \ldots, u_n = t_n \in A^*\}$. The theory $A^*$ is the closure under functional reflexivity of the basis $A^\sharp$. If $A$ is as in our first example, then $A^\sharp = \{2j = 0 : j > 0\}$.

**Completion.** On the other hand, if the ordering on leaves compares terms in some simplification ordering $\geqslant$, then the canonical basis will be the fully reduced set, as generated by (ground) completion: $\rho A = \{u = u\} \cup \{u = t : t = v \in A^*, t \gg v, v \text{ is not } u\}$. For our first example, $A^\sharp = \{2 = 0\}$. For another example, if $A = \{a = c, sa = b\}$ and $sa \gg sb \gg sc \gg a \gg b \gg c$, then $I(sa, b) > T(S(I(a, c)), I(sc, b))$, and hence $A^\sharp = \{a = c, sc = b\}$.

**Superposition.** If one distinguishes between $T$ steps based on the weight of the shared term $j$, making $T > I$ when $j$ is the smallest, and $T < I$ otherwise, then the canonical basis is also closed under paramodulation into the larger side of equations.

## 6 Derivations

Theorem proving with simplification entails two processes: **Expansion**, whereby any sound deductions (anything in $E^*$) may be added to the set of derived theorems; and **Contraction**, whereby any redundancies (anything in $\rho E$) may be removed.

A sequence of presentations $E_0 \rightsquigarrow E_1 \rightsquigarrow \cdots$ is called a *derivation*. Let $E_* = \cup_i E_i$. The *result* of the derivation is, as usual [Bachmair and Dershowitz, 1994], its *persisting* formulæ:

$$E_\infty \quad := \quad \liminf_{j \to \infty} E_j$$

We will say that a proof $p$ *persists* when $\Gamma p \subseteq E_\infty$. Thus, if a proof persists, so do its subproofs (by 6). By (16), we have $\Pi E_i \sqsupseteq \Pi E_*$.

**Definition 6** *A derivation $E_0 \rightsquigarrow E_1 \rightsquigarrow \cdots$ is* good *if $E_i \gtrsim E_{i+1}$ for all $i$.*

We are only interested in good derivations. From here on in, only good derivations will be considered. It is easy to see that:

**Lemma 4** *Derivations, the steps of which are expansions and contractions, are good.*

**Proposition 8** *If a derivation is good, then the limit supports the best proofs: $E_* \approx E_\infty$.*

**Proof.** One direction, namely $\Pi E_\infty \sqsupseteq \Pi E_*$, follows by (16) from the fact that $E_\infty \subseteq E_*$. To establish that $\Pi E_* \sqsupseteq \Pi E_\infty$, we show that $\mu \Pi E_* \sqsupseteq \Pi E_\infty$ and rely on (12). Suppose $p \in \mu \Pi E_*$. It follows from (5,9) that $\widehat{\Gamma p} \subseteq \Sigma p \subseteq \mu \Pi E_* \subseteq \mu \Pi E_*$. By goodness, each $a \in \Gamma p$ persists from some $E_i$ on. Hence, $\Gamma p \subseteq E_\infty$, and $p \in \Pi E_\infty$.

**Definition 7** *A good derivation is* fair *if $C(E_\infty) \sqsupset \Pi E_*$ where $C(E)$ is the set of* critical proof obligations*:*

$$C(E) \quad := \quad \{p \in \Pi E : p \notin \mu \Pi E^*, \; \forall q \lhd p.\; q \in \mu \Pi E^*\} \quad (24)$$

*It is* clean *if $\rho E_* \cap E_\infty = \emptyset$.*

Critical obligations are proofs that are not in normal form but all of whose proper subproofs are already in normal form. Fairness means that all persistent obligations are eventually "subsumed" by a strictly smaller proof.

**Lemma 5** *If a derivation is clean, then its limit is reduced.*

**Proof.** Suppose, on the contrary, that some $r \in \rho E_\infty \subseteq E_\infty \subseteq E_*$. Consider $\widehat{r}$, and compare it to a smaller proof $p \in \Pi E_\infty$. Let $A = \Gamma p \subseteq E_\infty \subseteq E_*$. Let $q \in \mu \Pi E_*$. Were $r \in \Gamma q$, then replacing $\widehat{r}$ as a subproof of $q$ with $p$, would by (7) result in a smaller proof than $q$. It follows that $r \in \rho E_*$, which contradicts cleanliness.

**Lemma 6** *If a derivation is fair, then its limit is complete.*

**Proof.** Any presentation $A$ is complete if $\Pi A \sqsupseteq \Pi A \cap \mu \Pi A^*$. since $a \in A^*$ implies $a \in \Delta(\Pi A \cap \mu \Pi A^*)$, whence completeness. Let $A = E_*$ be all formulæ proved at any stage in the derivation. We show that $A$ is complete in the above manner. Completeness of $E_\infty$ follows from Lemma 2. Consider any proof in $p \in \Pi A$ of $a$. Let $p_\infty \in \Pi E_\infty \subseteq \Pi A$ be the persisting proof of $a$, for which $p_\infty \leq p$ by the previous proposition. If $p_\infty \in \mu \Pi A^*$, we're done. Otherwise, $p_\infty$ has a minimal (with respect to $\trianglelefteq$) non-normal-form (possibly trivial) subproof $q$, all subproofs of which (persist and) are in normal form. By fairness, there is a proof $r \in \Pi A$ of the same theorem as $q$ such that $p_\infty \trianglerighteq q > r$. By (7), there is therefore a better proof $p' < p_\infty \leq p$. By induction, there is a $p'' \leq p'$ in both $\Pi A$ and $\mu \Pi A^*$, also proving $a$.

For example, suppose a proof ordering makes $\widehat{c} > \frac{\widehat{a}}{c}$ and $\frac{\widehat{c}}{a} > \widehat{a}$. Start with $E_0 = \{c\}$, and consider $\widehat{c}$. Were $\widehat{c}$ to persist, then by fairness a better proof would evolve, the better proof being $\frac{\widehat{a}}{c}$. If $\widehat{a}$ is in normal form, then $a \in E_\infty$ and both minimal proofs persist. Another example: $\mu \mathbb{P} = \{\widehat{a}, \widehat{c}, \frac{\widehat{a}}{c}\}$ and $E = \{a\}$, then $E \rightsquigarrow E \rightsquigarrow \cdots$ is fair, since $E_\infty = E$ and $C(E_\infty) = \emptyset$. The result is complete but unsaturated ($c$ is missing).

Together, these lemmata and Proposition 6 yield:

**Theorem 3** *If minimal proofs are unique and a derivation is fair and clean, then its limit is canonical.*

By (22), this also means that each $e \in E_\infty$ is its own ultimate proof $\widehat{e}$, so is not susceptible to contraction.

Returning to our main example, if projection $P$ is the most expensive type of inference, then no minimal proof includes it. And if proofs are compared in a simplification ordering (subproofs are always smaller than their superproofs), then minimal proofs will never have superfluous transitivity inferences of the form

$$\frac{u = t \quad t = t}{u = t} \; \mathbf{T}$$

Let $\gg$ be a total simplification-ordering of terms, let $P > I > T > S > Z$ in the precedence, let proofs be greater than terms, and compare proof trees in the corresponding total recursive path simplification-ordering. *Ground completion* is an inference mechanism consisting of the following inference rules:

**Deduce:** $\quad E \cup \{w = t[u]\} \quad \rightsquigarrow \quad E \cup \{w = t[v]\} \quad$ if $u = v \in E$ and $u \gg v$

**Delete:** $\quad E \cup \{t = t\} \quad \rightsquigarrow \quad E$

Furthermore, operationally, completion implements these inferences "fairly": No persistently enabled inference rule is ignored forever.

**Corollary 2 (Completeness of Completion)** *Ground completion results—at the limit—in the canonical, Church-Rosser basis:* $E_\infty = E_0^\sharp$.

**Proof.** Ground completion is good, since **Deduce** and **Delete** don't increase proofs ($\rightsquigarrow \subseteq \succsim$). In particular, $I(w, t[u]) > T(I(w, t[v]), S^n(I(u,v)))$ if $u \gg v$, since $t[u] \gg t[v]$ and $t[u] \geqq u \gg v$. Ground completion is fair and clean. For example, the critical obligation

$$\frac{w = t \quad t = v}{w = v} \ \mathbf{T}$$

when $t \gg w, v$, is resolved by **Deduce**. Also, since $T > S$, non-critical cases resolve naturally:

$$\frac{\dfrac{w=t}{fw=ft} \quad \dfrac{t=v}{ft=fv}}{fw=fv} \quad > \quad \frac{\dfrac{w=t \quad t=v}{w=v}}{fw=fv}$$

## 7 Discussion

We have suggested here that proof orderings, rather than formula orderings, take center stage in theorem proving with contraction (simplification and deletion of formulæ). Given a proof ordering that distinguishes "good proofs" from "bad proofs", it makes sense to define completeness of a set of formulæ as the claim that all theorems enjoy a smallest ("best") proof. Then an inference system is complete if it has the ability to generate all formulæ needed for such ideal proofs. Given a formula ordering, one can, of course, choose to compare proofs by simply comparing the multiset of their assumptions.[1]

The notion of "saturation" in theorem proving, in which superfluous deductions are not necessary for completeness, was suggested by Rusinowitch [1989, pp. 99–100] in the context of a Horn-clause resolution calculus. In our terminology: A presentation was said to be saturated when all inferrible formulæ are subsumed by formulæ in the presentation. This concept was refined by Bachmair and Ganzinger (see, most recently, [2001]) and by Nieuwenhuis and Rubio [2001, pp. 29–42]. They define saturation in terms of a more general kind of redundancy: An inference is redundant if its conclusion can be inferred from smaller formulæ; a presentation is saturated if every inference is redundant.

We propose alternate definitions of saturation and redundancy, defining both in terms of the proof ordering. This appears to be more flexible, since it allows small proofs to use large assumptions. The definition of redundancy in [Nieuwenhuis and Rubio, 2001] coincides with ours when proofs are measured first by their maximal assumption.

In [Bachmair and Dershowitz, 1994], a completion sequence is deemed fair if all persistent critical inferences are generated. In [Nieuwenhuis and Rubio, 2001, fn. 8], an inference sequence is held to be fair if all persistent inferences are either generated or become redundant. The definition of fairness propounded here combines the two ideas. But fairness only earns completeness, not saturation. (A stronger version of fairness is needed for saturation when the proof ordering is partial.) Our definition of critical obligations also allows one to incorporate "critical pair criteria" [Kapur *et al.*, 1988; Bachmair and Dershowitz, 1988].

---
[1] A well-founded proof ordering on $\mathbb{P}$ can induce a well-founded formula ordering on $\mathbb{A}$: Define $a \ggg c$ if there exist $p, q \in \mu\mathbb{P}$ such that $q \trianglerighteq p$, $\Delta p = a$, and $\Delta q = c$.

## Acknowledgment

I thank Claude Kirchner and my students for thinking and working together with me on these issues.